\documentstyle[prl,aps,multicol,epsfig,array]{revtex}

\newcommand \be{\begin{eqnarray}}
\newcommand \ee{\end{eqnarray}}

\begin{document}
\draft

\title{Streamer Branching rationalized by Conformal Mapping Techniques}

\author{Bernard Meulenbroek$^1$, Andrea Rocco$^1$ and Ute Ebert$^{1,2}$}

\address{
$^1$CWI, P.O.Box 94079, 1090 GB Amsterdam, The Netherlands,}
\address{
$^2$Dept.\ Physics, TU Eindhoven, 5600 MB Eindhoven, The Netherlands}

\date{\today}

\maketitle

\begin{abstract} 
Spontaneous branching of discharge channels is frequently
observed, but not well understood. We recently proposed a
new branching mechanism based on simulations of 
a simple continuous discharge model in high fields.
We here present analytical results for such streamers in the 
Lozansky-Firsov limit where they can be modelled as moving 
equipotential ionization fronts. This model can be analyzed
by conformal mapping techniques which allow the reduction 
of the dynamical problem to finite sets of nonlinear ordinary 
differential equations. Our solutions illustrate that branching 
is generic for the intricate head dynamics of streamers in the 
Lozansky-Firsov-limit.
\end{abstract}


\begin{multicols}{2}

When non-ionized matter is suddenly exposed to strong fields,
ionized regions can grow in the form of streamers. These are
ionized and electrically screened channels with rapidly propagating tips. 
The tip region is a very active impact ionization region due to 
the self-generated local field enhancement. Streamers appear
in early stages of atmospheric discharges like sparks or sprite discharges 
\cite{Pasko,PhysT}, they also play a prominent role in numerous 
technical processes. It is commonly observed that streamers branch 
spontaneously \cite{Williams02,Eddie02}.
But how this branching is precisely determined by
the underlying discharge physics, is essentially not known.
In recent work \cite{PRLMan,AndreaRapid}, we have
suggested a branching mechanism from first principles. 
This work drew some attention \cite{Nat,Focus}, since the proposed 
mechanism yields quantitative predictions for specific parameters,
and since it is qualitatively different from the older 
branching concept of the ``dielectric breakdown model'' 
\cite{DBM,DBM2,DBM3}. This older concept actually can be traced 
back to concepts of rare long-ranged (and hence stochastic) 
photo-ionization events probably first suggested in 1939 by Raether 
\cite{Raether}. Therefore, it came as a surprise that we predicted 
streamer branching in a fully deterministic model. Since our evidence 
for the phenomenon was mainly from numerical solutions together 
with a physical interpretation, the accuracy of our numerical scheme
was challenged \cite{KuliCom,Reply}. Furthermore, some authors
have argued previously \cite{PanchC,KuliR} that in a deterministic 
discharge model like ours, an initially convex streamer head never 
could become locally concave, and that hence the consecutive
branching of the discharge channel would be unphysical.

Therefore in the present paper, we investigate the issue by 
analytical means. We show that the convex-to-concave evolution of the streamer head with successive branching is generic for streamers in the 
Lozansky-Firsov limit \cite{PRLMan,AndreaRapid,Firsov}. We define the Lozansky-Firsov-limit 
as the stage of evolution where the streamer head is almost 
equipotential and surrounded by a thin electrostatic screening layer.
While in the original article \cite{Firsov}, only simple steady state 
solutions with parabolic head shape are discussed, we will show here
that a streamer in the Lozansky-Firsov limit actually
can exhibit a very rich head dynamics that includes spontaneous branching. 
Furthermore, our analytical solutions disprove the reasoning 
of \cite{PanchC} by explicit counterexamples. Our analytical 
methods are adapted from two fluid flow in Hele-Shaw cells 
\cite{rus,st,bensimon1,bensimon2,overview}. But our explicit and exact
solutions that amount to the evolution of ``bubbles'' in a dipole
field \cite{entov}, have not been reported in the hydrodynamics
literature either.

The relation between our previous numerical investigations
\cite{PRLMan,AndreaRapid} and our present analytical model is laid in 
two steps. First, numerical solutions show essentially the same evolution 
in the purely two-dimensional case as in the three-dimensional case 
with assumed cylinder geometry \cite{PRLMan,AndreaRapid}. 
Because there is an elegant analytical approach, we focus on the 
two-dimensional case. This has the additional advantage that such 
two-dimensional solutions rather directly apply to, e.g., discharges 
in Corbino discs \cite{Schoell}. Second, we use the following simplifying
approximations for a Lozansky-Firsov streamer: $(1)$ the interior of
the streamer is electrically completely screened; hence the electric potential 
$\varphi$ is constant; hence the ionization front
coincides with an equipotential line, $(2)$ the width of the 
screening layer around the ionized body is much smaller than
all other relevant length scales and in the present study it is actually 
neglected, $(3)$ the velocity of the ionization front {\bf v} is 
determined by the local electric field; in the simplest case to be
investigated here, it is simply taken to be proportional 
to the field at the boundary ${\bf v}=c\;\nabla\varphi$ with some
constant $c$ (for the validity 
of the approximation, cf.\ \cite{Ute,Sit}). Together with $\nabla^2\varphi=0$
in the non-ionized outer region and with fixed limiting values of the 
potential $\varphi$ far from the streamer, this defines a moving boundary 
problem for the interface between ionized and non-ionized region.
We assume the field far from the streamer to be constant as in our
simulations \cite{AndreaRapid}. Such a constant far field can be 
mimicked by placing the streamer between 
the two poles of an electric dipole where the distance between the poles
is much larger than the size of the streamer. 

When the electric field points into the $x$ direction and $y$ parametrizes 
the transversal direction, our two-dimensional Lozansky-Firsov streamer in 
free flight in a homogeneous electric field is approximated by:
\be  \label{bj1}
\nabla^2 \varphi(x,y) = 0, &\quad&\mbox{outside the streamer},\\ 
\label{bj3}\label{farfield}
-\nabla \varphi(x,y) \rightarrow E_0 \hat x, &\quad 
&\mbox{far outside the streamer}, \\ 
\label{bj2}
\varphi(x,y) = 0, &\quad&\mbox{inside the streamer},\\
{\bf v}_{\rm bound.} = c\;\nabla \varphi\big|_{\rm bound.}, 
&\quad &\mbox{velocity of the boundary}, \label{bj4}
\ee
where $\hat x$ is the unit vector in the $x$ direction, and we have
chosen the gauge such that the potential inside the
streamer vanishes. The asymptote (\ref{farfield}) implies that 
the total charge on the streamer vanishes; otherwise a contribution
$\propto 1/\sqrt{x^2+y^2}$ has to be added on the r.h.s.\ of Eq.\
(\ref{farfield}).

Similar moving boundary problems arise in Hele-Shaw flow 
of two fluids with a large viscosity contrast \cite{rus,st}:
Lozansky-Firsov streamers and viscous fingers on the present level 
of description can be identified immediately by equating the 
electric potential $\varphi$ with the fluid
pressure $p$ \cite{Ute,Sit}. To such problems, powerful conformal mapping 
methods \cite{bensimon1,bensimon2,overview} can be applied. 
Most work with this method is concerned with viscous fingers 
in a channel geometry, i.e., with boundary conditions on a lateral 
external boundary that cannot be realized in an electric system. 
A few authors also study 
air bubbles within a viscous fluid, or viscous droplets in air,
mostly under the action of flow fields generated by one source 
or one sink of pressure, i.e., by monopoles. 
On the other hand, the approximation (\ref{bj1})-(\ref{bj4}) describes
streamers in free space between two electrodes as in \cite{AndreaRapid}.
With the asymptote (\ref{bj3}), this is mathematically equivalent 
to air bubbles in a dipole field.
This case has not been studied 
in detail. It is known that any ellipse with the main axes
oriented parallel and perpendicular to the direction of the dipole
is a uniformly translating solution of this problem \cite{tanv1}.
The time dependent solutions of \cite{tanv2} do not apply
to streamers since the boundary condition on the moving interface
is different. Refs.\ \cite{entov} and \cite{nie} study how and when 
cusps in the interfaces of droplets and bubbles emerge when these 
are driven by multipole fields. But for bubbles in a dipole
field, again only the steady state ellipse solutions are given \cite{entov}.

In the present paper, we therefore apply conformal mapping methods
to the evolution of ``bubbles'' in a dipole field in a Hele-Shaw experiment
and proceed beyond the steady state ellipse solutions. We identify the 
general structure of time dependent solutions of (\ref{bj1})-(\ref{bj4}).
The analytically derived exact solutions show how a streamer head can become 
flatter, concave and branch as observed numerically \cite{PRLMan,AndreaRapid}.
Rather than a pole decomposition \cite{bensimon1}, 
we derive a decomposition into Fourier modes of the circle and calculate
an equation for the non-linear dynamical coupling of their amplitudes.

In detail, this is done in the following steps:
\\ 
$(i)$ The spatial coordinates are expressed by the complex coordinate 
$z=x+iy$. According to standard complex analysis, finding a real 
harmonic function $\varphi(x,y)$ solving the Laplace equation (\ref{bj1})
in a given region is equivalent to finding a complex function 
$\Phi(z)$ that is analytical in the same region and has real part 
${\rm Re}\;\Phi(z)=\varphi(x,y)$.
\\
$(ii)$ A conformal map from the interior of the unit circle to the exterior 
of the streamer or ``bubble'' is constructed. Including the point at
infinity, the region outside the streamer is simply connected
and Riemann's mapping theorem applies; therefore the mapping exists.
Since the boundary moves, the mapping is time dependent; we denote it with
$z=f_t(\omega)$ where $\omega$ parametrizes the interior of the unit circle
$|\omega|<1$. The complete map can be composed from a conformal map 
$\zeta=h_t(\omega)$ that deforms the unit disc continuously, followed 
by the inversion $z=1/\zeta$. Since $h_t(\omega)$ is conformal on the 
unit disc, it is analytical and has a single zero 
which we choose to be at $\omega=0$. Therefore $f_t(\omega)=1/h_t(\omega)$ 
has a single pole $\propto \omega^{-1}$ and is otherwise analytical. 
Rather than a pole decomposition \cite{bensimon1}, 
we choose a Laurent expansion for $f_t(\omega)$
\be
\label{map}
x+iy=z=f_t(\omega)=\sum_{k=-1}^\infty a_k(t)\;\omega^k.
\ee
This expansion allows us to identify the exact dynamical solutions 
(\ref{solN}) below. Taking $a_{-1}(t)$ as a real positive number 
makes the mapping unique, again according to Riemann's mapping theorem. 
\\
$(iii)$ Now the potential $\hat\Phi(\omega)$ on the unit disc can
be calculated explicitly. Since $f_t(\omega)$ is a conformal mapping, 
the function $\Phi(z)$ is analytical if and only if 
the function $\hat \Phi(\omega)=\Phi(f_t(\omega))$ is analytical. 
The asymptote of $\hat\Phi(\omega)$ for $\omega\to0$ is determined
by (\ref{farfield}) and (\ref{map}): for $|x|,|z|\to\infty$, we 
have $\varphi(x,y)\to-E_0x$, hence $\Phi(z)\to-E_0z$, and therefore
with (\ref{map}): $\hat\Phi(\omega)\to -E_0a_{-1}(t)/\omega$ for $\omega\to0$.
This means that the pole of $\hat\Phi(\omega)$ at the origin of the unit disc 
$\omega =0 $ corresponds to the dipole of $\Phi(z)$ at $z\to\pm\infty$.
This dipole generates the field and the interfacial motion. In the
remainder of the unit disc, there are no sources or sinks of potential,
hence $\hat\Phi$ is analytical there. Furthermore, at the boundary 
of the streamer, we have $\varphi=0$ from (\ref{bj2}) or ${\rm Re}\;\Phi=0$, 
resp. The boundary of the streamer maps onto the unit circle, 
so ${\rm Re}\;\hat\Phi(\omega)=0$ for $|\omega|=1$. Using the asymptotics
at $\omega\to0$ and analyticity in the remaining region, the unique and
exact solution for the potential is 
\be \label{bern8}
\hat \Phi(\omega) = E_0 a_{-1}(t)\left(\omega - \frac{1}{\omega}\right).
\ee
$(iv)$ The velocity ${\bf v}_{\rm bound.}=c\nabla\phi$ (\ref{bj4}) 
determines the motion of the interface. 
This interface is the time dependent map $f_t(\omega)$ of the unit circle 
$\omega=e^{i\alpha}$ parametrized by the angle $\alpha\in[0,2\pi)$.
Therefore Eq.\ (\ref{bj4}) determines the dynamics of 
$f_t\left(e^{i\alpha}\right)$. According to \cite{rus,bensimon1}, it is
\begin{equation}
{\rm Re}\;\Big[\!-i \partial_\alpha f_t^*\big(e^{i\alpha}\big) \;
\partial_t f_t\big(e^{i\alpha}\big) \Big] = 
c\;{\rm Re}\;\Big[i\partial_\alpha\hat\Phi\big(e^{i\alpha}\big)\Big].
\label{boundary}
\end{equation}

The problem (\ref{bj1})-(\ref{bj4}) is symmetric under reflection 
on the $x$-axis. We assume the solutions to have 
the same symmetry. Therefore all $a_k(t)$ have to be real.
The position $(x,y)(\alpha,t)$ of the point of the interface labelled
by the angle $\alpha$ at time $t$ can be read directly from 
the Laurent expansion (\ref{map}) by inserting $\omega=e^{i\alpha}$;
it has essentially the form of a Fourier expansion of the unit circle
where the circle and its position are created by the modes $k=-1$ and 0:
\begin{eqnarray}
\label{inter}
&&x(\alpha,t)=\!\sum_{k=-1}^\infty\! a_k(t)\;\cos k\alpha~~,~~
y(\alpha,t)=\!\sum_{k=-1}^\infty\! a_k(t)\;\sin k\alpha,
\nonumber\\
&&~~~a_k(t) ~\mbox{real}~~,~~ a_{-1}(t)>0~~~,~~~0\le\alpha<2\pi~.
\end{eqnarray}

Substituting the mapping function (\ref{map}) and the potential
(\ref{bern8}) into the equation of motion for the mapping (\ref{boundary}),
and assuming the $a_k(t)$ to be real, we obtain for the evolution 
of the amplitudes $a_k(t)$:
\be \label{eq}
&&\sum_{k,k'=-1}^{\infty} k'a_{k'}(t)\;\partial_ta_k(t)
\;\cos\big((k-k')\alpha\big)
\nonumber\\
&&\qquad\qquad = 2E_0c\;a_{-1}(t) \;\cos\alpha.
\ee
A closer investigation shows that this equation has an important 
property: suppose that the streamer boundary can be written initially 
as a finite series $\sum_{k=-1}^N a_k(0)\; e^{ik\alpha}$, $a_N(0)\ne0$. 
Then at all times $t$, the interface is described by the same finite 
number of modes
\be
\label{solN}
z(\alpha,t)=\sum_{k=-1}^N a_k(t)\;e^{ik\alpha},
\ee 
i.e., the $a_k(t)$ with $k>N$ stay identical to zero at all times $t>0$.
Sorting the terms in (\ref{eq}) by coefficients of $\cos k\alpha$, 
the equation can be recast into $N+2$ ordinary differential equations 
for the $N+2$ functions $a_k(t)$
\be \label{eqN}
\lefteqn{
\sum_{k=-1}^{N-m}\Big[(k+m)\;a_{k+m}\;\partial_ta_k
+k\;a_k\;\partial_ta_{k+m}\Big]}
\nonumber\\
&&~~=2E_0c\;a_{-1}\;\delta_{m,1}~~~~\mbox{ for } m=0,\ldots,N+1,
\ee
where $\delta_{m,1}$ is the Kronecker symbol. 
Eq.~(\ref{eqN}) is equivalent to a matrix equation of the form
${\bf A}(\{a_k(t)\}) \cdot \partial_t
\Big(a_{-1}(t), \ldots, a_N(t)\Big)=
\Big(0, 2E_0c a_{-1}(t), 0, \ldots, 0\Big)$,
where the matrix ${\bf A}$ depends linearly on the $\{a_k(t)\}$.

Eqs.~(\ref{solN}) and (\ref{eqN}) identify large classes of analytical 
solutions with arbitrary fixed $N$. These solutions reduce the dynamical 
moving boundary problem in two spatial dimensions of Eqs.\ 
(\ref{bj1})--(\ref{bj4})
exactly to a finite set of ordinary differential equations for 
the nonlinear coupling
of the amplitudes $a_k(t)$ of modes $e^{ik\alpha}$, $0\le\alpha<2\pi$. 
These equations
are easy to integrate numerically or for small $N$ even analytically. 
We will 
use this form to discuss now generic solutions of Eqs.\ 
(\ref{bj1})--(\ref{bj4}) 
as the simplest approximation of a streamer in the Lozansky-Firsov limit.

First, it is now easy to reproduce the uniformly pro\-pa\-gating 
ellipse solutions of \cite{entov,tanv1} as the solutions with $N=1$: 
for $|a_1|\ne|a_{-1}|$, the equations reduce to 
$\partial_ta_{-1}=0=\partial_ta_1$ and 
$\partial_ta_0=2E_0c\;a_{-1}/(a_1-a_{-1})$.
These solutions correspond to ellipses whose principal radii are 
oriented along the axes. These radii maintain their
values $r_{x,y}=a_{-1}\pm a_1$ (assuming $a_{-1}>a_1>0$) and move 
with constant velocity $v_{\rm ellipse} = -E_0c\;(r_x+r_y)/r_y$.
The Lozansky-Firsov-parabola can be understood as limit cases 
of such uniformly propagating ellipses.

In contrast to $N\le1$, all solutions with $N\ge2$ have nontrivial 
dynamics. It can be tracked
by integrating the $N+2$ ordinary differential equations (\ref{eqN}) 
numerically and then plotting the boundaries (\ref{solN}) at 
consecutive times. Examples of such dynamics are shown in the figures.

\begin{figure}
\includegraphics[height=5.6cm]{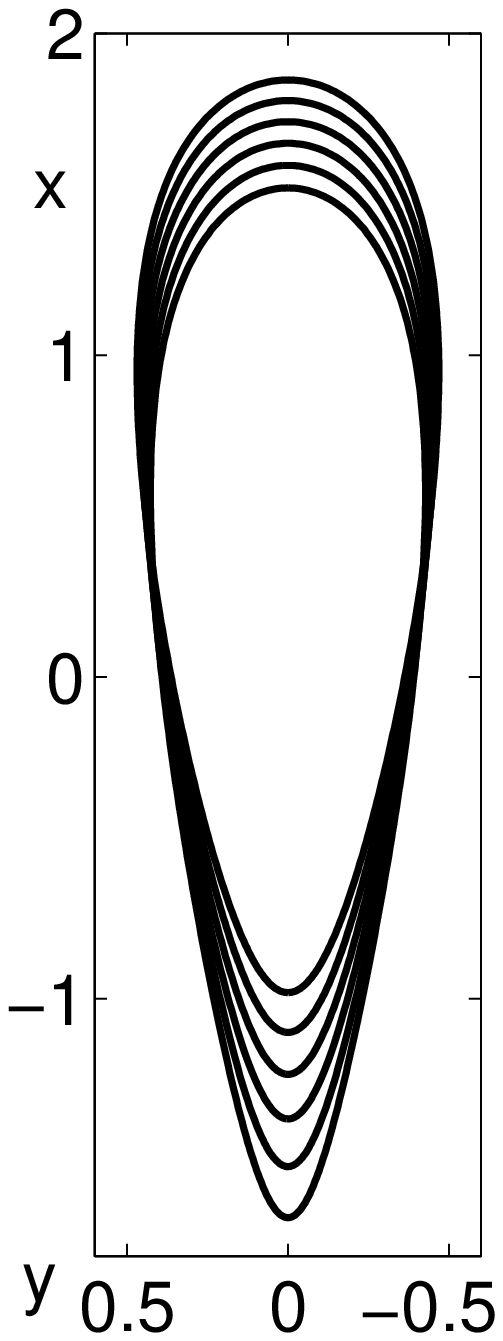}
\includegraphics[height=5.6cm]{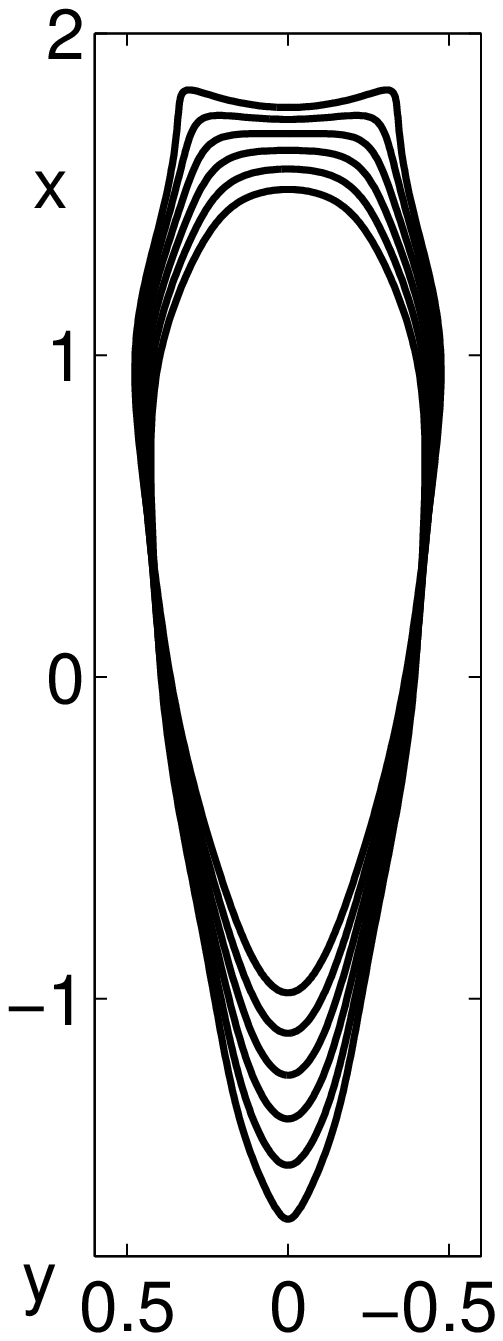}
\includegraphics[height=5.6cm]{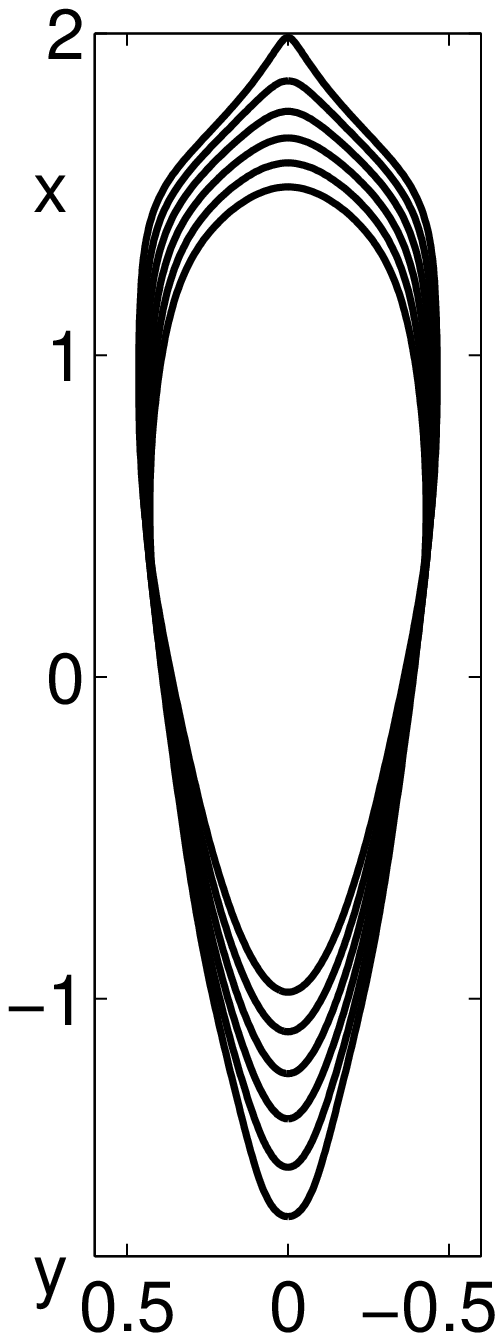}
\includegraphics[height=5.6cm]{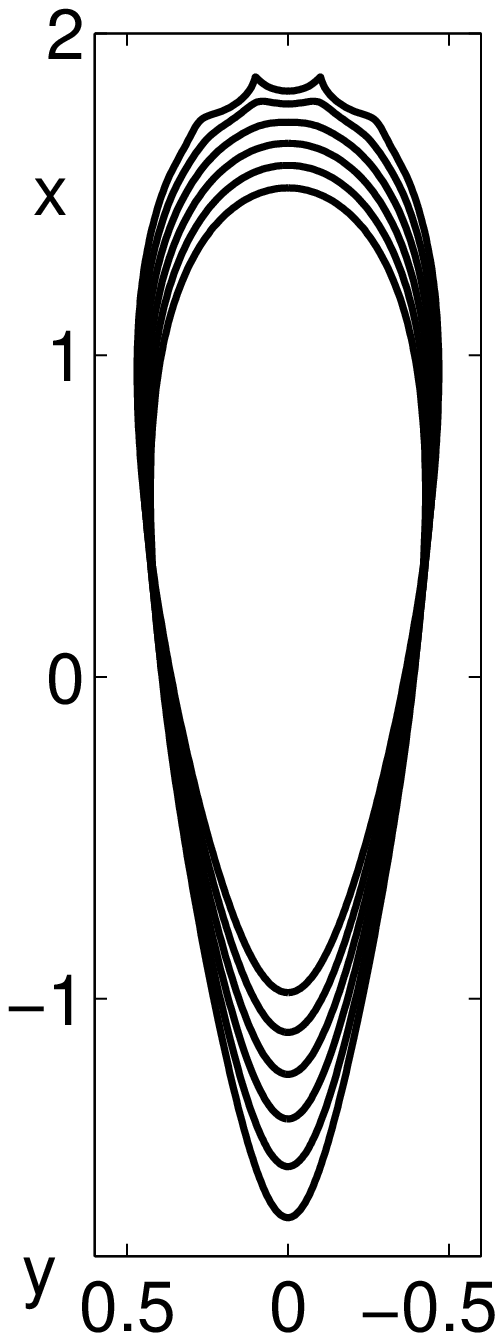}

\vspace{-0.1cm} 

\centerline{
\includegraphics[height=2.5cm]{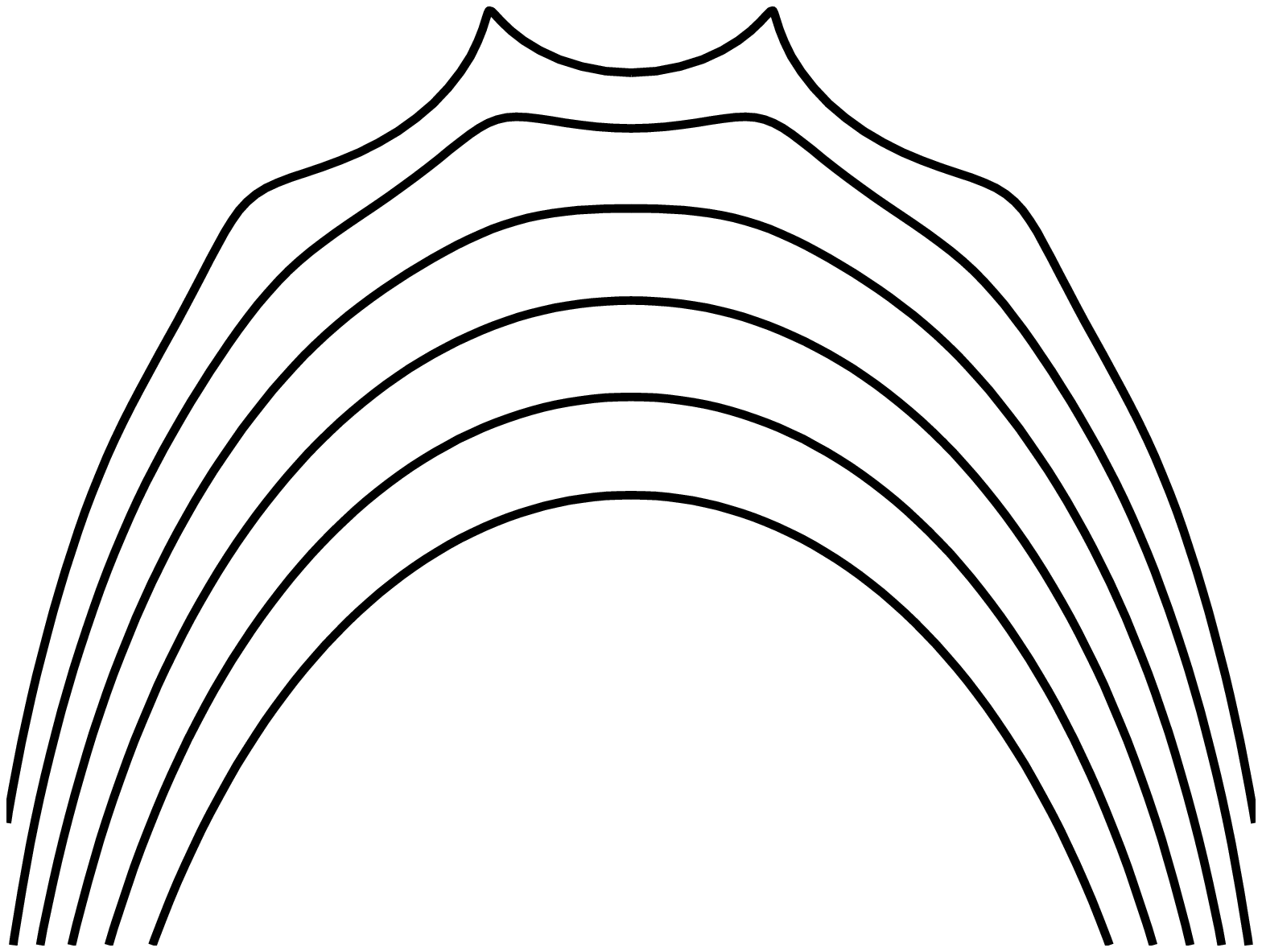}
}
\caption{Upper panel: evolution of the interface  in equal time steps 
up to time $t=0.1/(E_0c)$ with initial condition 
\newline
a) $z_0(\alpha,0)=e^{-i\alpha}+0.6\cdot e^{i\alpha}-0.08\cdot e^{2i\alpha}$, 
\newline
b) $z(\alpha,0)=z_0(\alpha,0)-5\cdot 10^{-3}\cdot e^{8i\alpha}$
\newline
c) $z(\alpha,0)=z_0(\alpha,0)+3\cdot 10^{-3}\cdot e^{8i\alpha}$
\newline
d) $z(\alpha,0)=z_0(\alpha,0)-4.5\cdot 10^{-7}\cdot e^{30i\alpha}$,
\newline
and lower panel: zoom into the unstable head of Fig.~d.
}
\end{figure}

Fig.\ 1 shows four cases of the upward motion of a conically shaped 
streamer in equal time steps. The initial conditions are almost
identical. On the leftmost figure, an ellipse is corrected only by 
a mode $e^{2i\alpha}$ to create the conical shape. 
This shape with $N=2$ eventually develops a concave tip, but only 
after much longer times than shown in the figure. In the other figures 
this conical shape is perturbed initially by a minor perturbation 
with wavenumber 8 or 30, corresponding to $N=8$ and 30 in (\ref{solN}) 
and (\ref{eqN}). The amplitude of the perturbation is chosen such
that a cusp develops at time $0.1/(E_0c)$. Depending on the sign of the
amplitude, the cusp develops on or off axis --- where we stress that
we are not interested in the cusp itself, but in the earlier stages
of evolution. Note that our reduction of the moving boundary problem 
to the set of ordinary differential equations (\ref{eqN}) assures 
that the evolving shape 
is a true solution of the problem (\ref{bj1})--(\ref{bj4}). Figs.\ $1b$ 
and $1d$ demonstrate that spontaneous branching is a possible solution.

\vspace{-0.3cm}

\begin{figure}
\includegraphics[height=5.0cm]{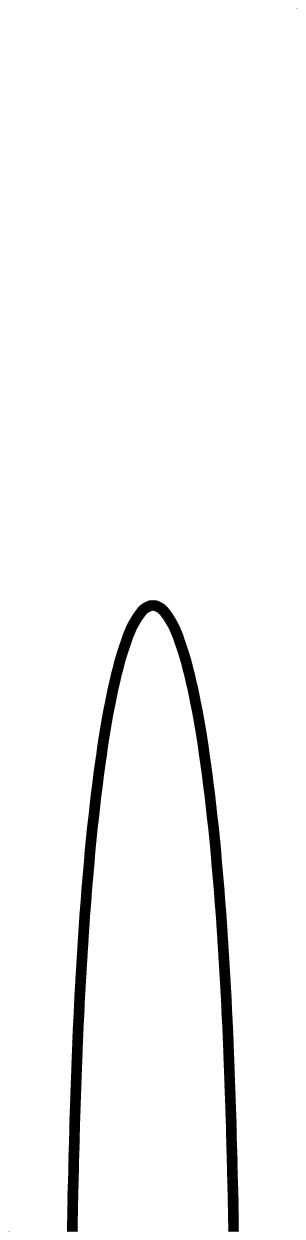} 
\includegraphics[height=5.0cm]{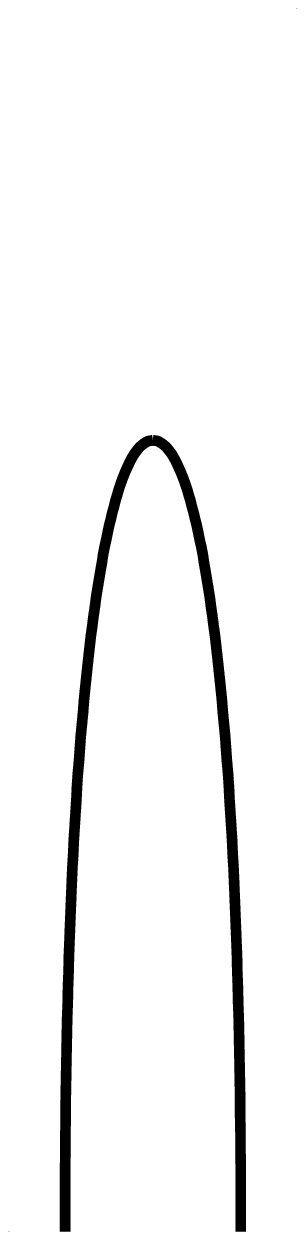}
\includegraphics[height=5.0cm]{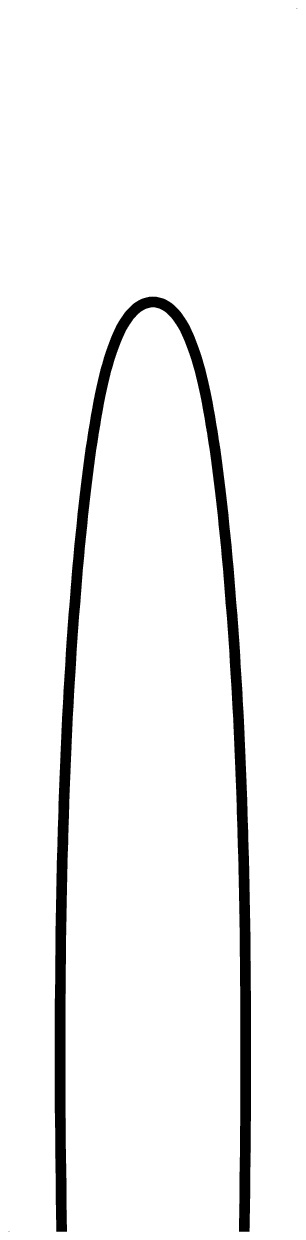}
\includegraphics[height=5.0cm]{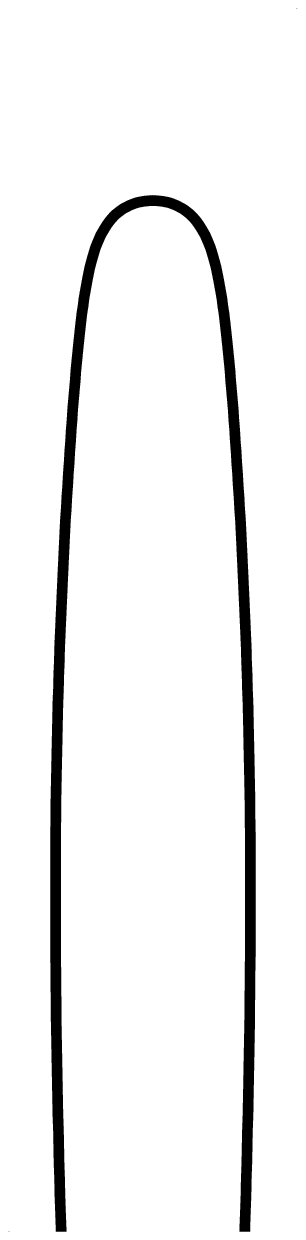}
\includegraphics[height=5.0cm]{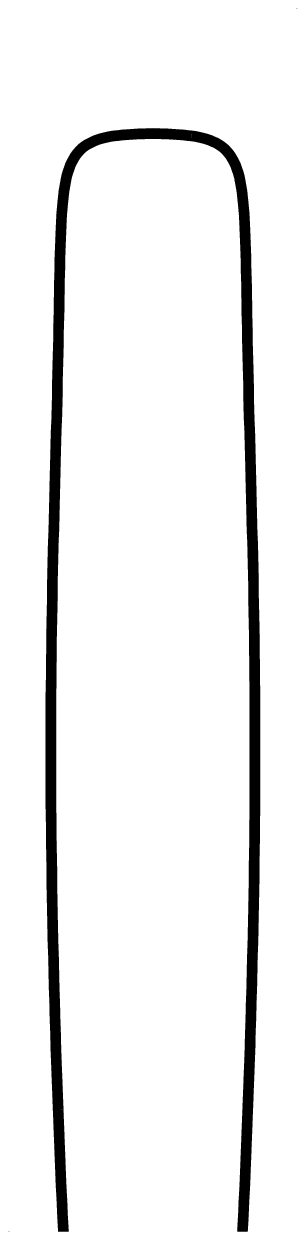}
\includegraphics[height=5.0cm]{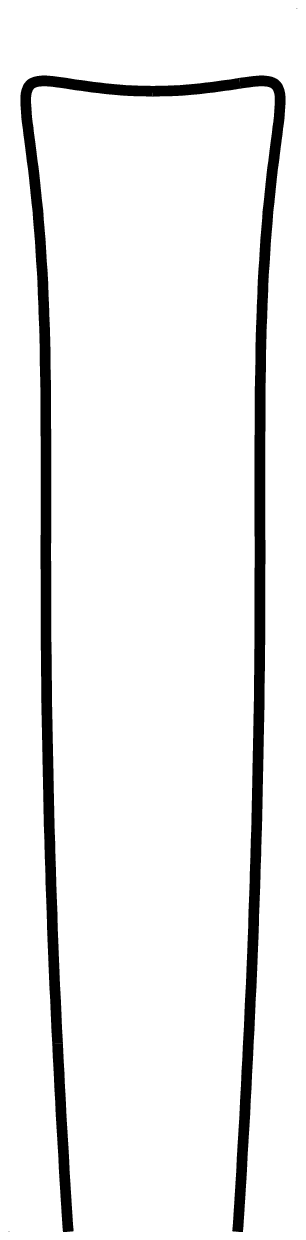}
\caption{Evolution of the tip of an elongated ``streamer'' in equal 
time steps up to time $t=0.1/(E_0c)$; initial condition 
$z(\alpha,0)=e^{-i\alpha}+0.9\cdot e^{i\alpha}
-0.03\cdot e^{2i\alpha}-1.2\cdot 10^{-5}\cdot e^{12i\alpha}$.
}
\end{figure}

In Fig.\ 2 the ionized body is longer stretched and only the
tip is shown, again at 6 equidistant time steps. The streamer
becomes slower when the head becomes flatter, since the electric
field then diminishes. Eventually, the head becomes concave and ``branches''.
 
In summary, the solutions of the moving boundary problem 
(\ref{bj1})--(\ref{bj4}) demonstrate
the onset of branching within a purely deterministic model. 
They show a high sensitivity to minor deviations
of the initial conditions. A streamer in the Lozansky-Firsov-limit 
is therefore also very sensitive to physical perturbations
during the evolution, and simulations in this limit 
are just as sensitive to small numerical errors. 
But perturbations during the evolution are not necessary for branching.

Our analysis applies to streamers in the Lozansky-Firsov-limit, 
i.e., to almost equipotential streamers that are surrounded by 
a very thin electrical screening layer. This limit is approached 
in our previous simulations \cite{PRLMan,AndreaRapid}.

These results raise the following questions that are presently 
under investigation:
1) When does a streamer reach this Lozansky-Firsov-limit
that then generically leads to branching?
2) The formation of cusps should be suppressed by some microscopic
stabilization mechanism. Is the electric screening length 
discussed in \cite{PRLMan,prep} sufficient to supply this mechanism?
3) If this stabilization is taken into account, can an interfacial
model reproduce numerical and physical streamer branching quantitavely?
4) How can the motion of the back end of the streamer be modelled
appropriately (rather than assuming the velocity law ${\bf v}\propto
\nabla\varphi$ (\ref{bj4}) everywhere)? How can it be incorporated into
the present analysis?

{\bf Acknowledgment:} B.M.\ was supported by CWI Amsterdam,
and A.R.\ by the Dutch Research School CPS, 
the Dutch Physics Funding Agency FOM and CWI.


\end{multicols}

\end{document}